\pdfoutput=1
\documentclass[12pt,a4paper]{article}

\usepackage{ifthen} 
\newboolean{pdflatex}
\setboolean{pdflatex}{true} 

\newboolean{articletitles}
\setboolean{articletitles}{true} 

\newboolean{uprightparticles}
\setboolean{uprightparticles}{false} 

\def\paperauthors{LHCb collaboration} 
\def\paperasciititle{Measurement of the mass difference and relative production rate of the Omegab- and Xib- baryons} 
\def\papertitle{Measurement of the mass difference and relative production rate \\ of the ${\mathchar"710A}^-_{\ensuremath{b}}$ and ${\mathchar"7104}^-_{\ensuremath{b}}$ baryons} 
\def\paperkeywords{{High Energy Physics}, {LHCb}} 
\def\papercopyright{\the\year\ CERN for the benefit of the LHCb collaboration} 
\def\paperlicence{CC BY 4.0 licence}
\def\paperlicenceurl{https://creativecommons.org/licenses/by/4.0/}

\def\LbToLJPsi{\decay{\Lb}{\jpsi\Lz}}
\def\XibToXiJPsi{\decay{\Xires^-_b}{\jpsi\Xires^-}}

\def\OmegabToOmegaJPsi{\decay{\Omegares^-_b}{\jpsi\Omegares^- }}

\def\splot{\ensuremath{_s\mathcal{P}lot}\xspace}

\def\pp{\ensuremath{\proton\proton}\xspace}

\usepackage[top=1in, bottom=1.25in, left=1in, right=1in]{geometry}

\usepackage{microtype}
\usepackage{lineno}  
\usepackage{xspace} 
\usepackage{caption} 

\usepackage{graphicx}  
\usepackage{color}
\usepackage{colortbl}
\graphicspath{{./figs/}} 

\usepackage{amsmath} 
\usepackage{amssymb}
\usepackage{booktabs}
\usepackage{amsfonts}
\usepackage{upgreek} 

\newcommand*\patchAmsMathEnvironmentForLineno[1]{%
\expandafter\let\csname old#1\expandafter\endcsname\csname #1\endcsname
\expandafter\let\csname oldend#1\expandafter\endcsname\csname
end#1\endcsname
 \renewenvironment{#1}%
   {\linenomath\csname old#1\endcsname}%
   {\csname oldend#1\endcsname\endlinenomath}%
}
\newcommand*\patchBothAmsMathEnvironmentsForLineno[1]{%
  \patchAmsMathEnvironmentForLineno{#1}%
  \patchAmsMathEnvironmentForLineno{#1*}%
}
\AtBeginDocument{%
\patchBothAmsMathEnvironmentsForLineno{equation}%
\patchBothAmsMathEnvironmentsForLineno{align}%
\patchBothAmsMathEnvironmentsForLineno{flalign}%
\patchBothAmsMathEnvironmentsForLineno{alignat}%
\patchBothAmsMathEnvironmentsForLineno{gather}%
\patchBothAmsMathEnvironmentsForLineno{multline}%
\patchBothAmsMathEnvironmentsForLineno{eqnarray}%
}

\usepackage{hyperxmp}

\usepackage[pdftex,
            pdfauthor={\paperauthors},
            pdftitle={\paperasciititle},
            pdfkeywords={\paperkeywords},
            pdfcopyright={Copyright (C) \papercopyright},
            pdflicenseurl={\paperlicenceurl}]{hyperref}

\usepackage[colorinlistoftodos,textsize=scriptsize]{todonotes}

\usepackage[bottom,flushmargin,hang,multiple]{footmisc}

\usepackage[all]{hypcap} 

\usepackage{xspace} 
\usepackage{upgreek}

\def\lhcb   {\mbox{LHCb}\xspace}

\def\tevatron {Tevatron\xspace}

\def\MagUp {\mbox{\em Mag\kern -0.05em Up}\xspace}

\ifthenelse{\boolean{uprightparticles}}%
{

 \def\Pmu         {\ensuremath{\upmu}\xspace}

 \def\Ppi         {\ensuremath{\uppi}\xspace}

 \def\Ppsi        {\ensuremath{\uppsi}\xspace}

 \def\PDelta      {\ensuremath{\Delta}\xspace}                 
 \def\PXi         {\ensuremath{\Xi}\xspace}                 
 \def\PLambda     {\ensuremath{\Lambda}\xspace}                 
 \def\PSigma      {\ensuremath{\Sigma}\xspace}                 
 \def\POmega      {\ensuremath{\Omega}\xspace}                 
 \def\PUpsilon    {\ensuremath{\Upsilon}\xspace}
 \let\oldPi\Pi
 \def\PPi         {\ensuremath{\oldPi}\xspace}

 \def\PB      {\ensuremath{\mathrm{B}}\xspace}                 
                  
 \def\PD      {\ensuremath{\mathrm{D}}\xspace}

 \def\PJ      {\ensuremath{\mathrm{J}}\xspace}                 
 \def\PK      {\ensuremath{\mathrm{K}}\xspace}

 \def\Pb      {\ensuremath{\mathrm{b}}\xspace}                 
 \def\Pc      {\ensuremath{\mathrm{c}}\xspace}                 
 \def\Pd      {\ensuremath{\mathrm{d}}\xspace}

 \def\Pi      {\ensuremath{\mathrm{i}}\xspace}

 \def\Pp      {\ensuremath{\mathrm{p}}\xspace}

 \def\Ps      {\ensuremath{\mathrm{s}}\xspace}                 
                  
 \def\Pu      {\ensuremath{\mathrm{u}}\xspace}

 \def\thebaroffset{0.0em}
}
{

 \def\Pmu         {\ensuremath{\mu}\xspace}

 \def\Ppi         {\ensuremath{\pi}\xspace}

 \def\Ppsi        {\ensuremath{\psi}\xspace}                 
                  
 \mathchardef\PDelta="7101
 \mathchardef\PXi="7104
 \mathchardef\PLambda="7103
 \mathchardef\PSigma="7106
 \mathchardef\POmega="710A
 \mathchardef\PUpsilon="7107
 \mathchardef\PPi="7105
                  
 \def\PB      {\ensuremath{B}\xspace}                 
                  
 \def\PD      {\ensuremath{D}\xspace}

 \def\PJ      {\ensuremath{J}\xspace}                 
 \def\PK      {\ensuremath{K}\xspace}

 \def\Pb      {\ensuremath{b}\xspace}                 
 \def\Pc      {\ensuremath{c}\xspace}                 
 \def\Pd      {\ensuremath{d}\xspace}

 \def\Pi      {\ensuremath{i}\xspace}

 \def\Pp      {\ensuremath{p}\xspace}

 \def\Ps      {\ensuremath{s}\xspace}                 
                  
 \def\Pu      {\ensuremath{u}\xspace}

 \def\thebaroffset{0.18em}
}
\newcommand{\offsetoverline}[2][\thebaroffset]{\kern #1\overline{\kern -#1 #2}}%

\makeatletter
\ifcase \@ptsize \relax
  \newcommand{\miniscule}{\@setfontsize\miniscule{4}{5}}
\or
  \newcommand{\miniscule}{\@setfontsize\miniscule{5}{6}}
\or
  \newcommand{\miniscule}{\@setfontsize\miniscule{5}{6}}
\fi
\makeatother

\DeclareRobustCommand{\optbar}[1]{\shortstack{{\miniscule (\rule[.5ex]{1.25em}{.18mm})}
  \\ [-.7ex] $#1$}}

\def\mumu       {{\ensuremath{\Pmu^+\Pmu^-}}\xspace}

\def\uquark    {{\ensuremath{\Pu}}\xspace}

\def\dquark    {{\ensuremath{\Pd}}\xspace}

\def\squark    {{\ensuremath{\Ps}}\xspace}

\def\cquark    {{\ensuremath{\Pc}}\xspace}

\def\bquark    {{\ensuremath{\Pb}}\xspace}

\def\pion   {{\ensuremath{\Ppi}}\xspace}
\def\piz    {{\ensuremath{\pion^0}}\xspace}

\def\pim    {{\ensuremath{\pion^-}}\xspace}

\def\kaon    {{\ensuremath{\PK}}\xspace}

\def\KorKbar {\kern \thebaroffset\optbar{\kern -\thebaroffset \PK}{}\xspace}

\def\Kp      {{\ensuremath{\kaon^+}}\xspace}
\def\Km      {{\ensuremath{\kaon^-}}\xspace}

\def\KS      {{\ensuremath{\kaon^0_{\mathrm{S}}}}\xspace}

\def\D       {{\ensuremath{\PD}}\xspace}

\def\DorDbar {\kern \thebaroffset\optbar{\kern -\thebaroffset \PD}\xspace}

\def\Dp      {{\ensuremath{\D^+}}\xspace}
\def\Dm      {{\ensuremath{\D^-}}\xspace}

\def\DpDm    {\ensuremath{\Dp {\kern -0.16em \Dm}}\xspace}

\def\B       {{\ensuremath{\PB}}\xspace}

\def\BorBbar {\kern \thebaroffset\optbar{\kern -\thebaroffset \PB}\xspace}

\def\Bd      {{\ensuremath{\B^0}}\xspace}

\def\BdorBdbar {\kern \thebaroffset\optbar{\kern -\thebaroffset \Bd}\xspace}
\def\Bu      {{\ensuremath{\B^+}}\xspace}

\def\Bp      {{\ensuremath{\Bu}}\xspace}

\def\Bs      {{\ensuremath{\B^0_\squark}}\xspace}

\def\BsorBsbar {\kern \thebaroffset\optbar{\kern -\thebaroffset \Bs}\xspace}

\def\jpsi     {{\ensuremath{{\PJ\mskip -3mu/\mskip -2mu\Ppsi}}}\xspace}

\def\Y#1S{\ensuremath{\PUpsilon{(#1S)}}\xspace}

\def\proton      {{\ensuremath{\Pp}}\xspace}

\def\Lz          {{\ensuremath{\PLambda}}\xspace}

\def\LorLbar     {\kern \thebaroffset\optbar{\kern -\thebaroffset \PLambda}\xspace}
\def\Lambdares   {{\ensuremath{\PLambda}}\xspace}

\def\Sigmares    {{\ensuremath{\PSigma}}\xspace}

\def\Xires       {{\ensuremath{\PXi}}\xspace}

\def\Xim         {{\ensuremath{\Xires^-}}\xspace}

\def\Omegares    {{\ensuremath{\POmega}}\xspace}

\def\Omegam      {{\ensuremath{\Omegares^-}}\xspace}

\def\Xicz        {{\ensuremath{\Xires^0_\cquark}}\xspace}
\def\Xicp        {{\ensuremath{\Xires^+_\cquark}}\xspace}

\def\Omegac      {{\ensuremath{\Omegares^0_\cquark}}\xspace}

\def\Lb           {{\ensuremath{\Lz^0_\bquark}}\xspace}

\def\Xib          {{\ensuremath{\Xires_\bquark}}\xspace}
\def\Xibz         {{\ensuremath{\Xires^0_\bquark}}\xspace}
\def\Xibm         {{\ensuremath{\Xires^-_\bquark}}\xspace}

\def\Omegab       {{\ensuremath{\Omegares^-_\bquark}}\xspace}

\newcommand{\decay}[2]{\ensuremath{#1\!\to #2}\xspace} 

\def\to                 {\ensuremath{\rightarrow}\xspace}

\def\AT#1     {\ensuremath{A_{\mathrm{T}}^{#1}}\xspace}           

\def\C#1      {\ensuremath{\mathcal{C}_{#1}}\xspace}                       
\def\Cp#1     {\ensuremath{\mathcal{C}_{#1}^{'}}\xspace}                    
\def\Ceff#1   {\ensuremath{\mathcal{C}_{#1}^{\mathrm{(eff)}}}\xspace}        
\def\Cpeff#1  {\ensuremath{\mathcal{C}_{#1}^{'\mathrm{(eff)}}}\xspace}       
\def\Ope#1    {\ensuremath{\mathcal{O}_{#1}}\xspace}                       
\def\Opep#1   {\ensuremath{\mathcal{O}_{#1}^{'}}\xspace}                    

\newcommand{\nospaceunit}[1]{\ensuremath{\text{#1}}}       
\newcommand{\aunit}[1]{\ensuremath{\text{\,#1}}}       

\newcommand{\tev}{\aunit{Te\kern -0.1em V}\xspace}
\newcommand{\gev}{\aunit{Ge\kern -0.1em V}\xspace}
\newcommand{\mev}{\aunit{Me\kern -0.1em V}\xspace}
\newcommand{\kev}{\aunit{ke\kern -0.1em V}\xspace}
\newcommand{\ev}{\aunit{e\kern -0.1em V}\xspace}
 
\newcommand{\mevc}{\ensuremath{\aunit{Me\kern -0.1em V\!/}c}\xspace}
\newcommand{\gevc}{\ensuremath{\aunit{Ge\kern -0.1em V\!/}c}\xspace}
\newcommand{\mevcc}{\ensuremath{\aunit{Me\kern -0.1em V\!/}c^2}\xspace}
\newcommand{\gevcc}{\ensuremath{\aunit{Ge\kern -0.1em V\!/}c^2}\xspace}

\def\mum  {\ensuremath{\,\upmu\nospaceunit{m}}\xspace}

\def\fb   {\ensuremath{\aunit{fb}}\xspace}
\def\invfb   {\ensuremath{\fb^{-1}}\xspace}

\def\ps   {\ensuremath{\aunit{ps}}\xspace}

\newcommand{\stat}{\aunit{(stat)}\xspace}
\newcommand{\syst}{\aunit{(syst)}\xspace}

\def\gsim{{~\raise.15em\hbox{$>$}\kern-.85em
          \lower.35em\hbox{$\sim$}~}\xspace}
\def\lsim{{~\raise.15em\hbox{$<$}\kern-.85em
          \lower.35em\hbox{$\sim$}~}\xspace}

\def\pt         {\ensuremath{p_{\mathrm{T}}}\xspace}

\def\ptot       {\ensuremath{p}\xspace}

\def\evtgen     {\mbox{\textsc{EvtGen}}\xspace}

\def\geant      {\mbox{\textsc{Geant4}}\xspace}

\def\photos     {\mbox{\textsc{Photos}}\xspace}

\def\pythia     {\mbox{\textsc{Pythia}}\xspace}

\def\tell1  {TELL1\xspace}
\def\ukl1   {UKL1\xspace}

\newcommand{\lhcborcid}[1]{\href{https://orcid.org/#1}{\hspace*{0.1em}\raisebox{-0.45ex}{\includegraphics[width=1em]{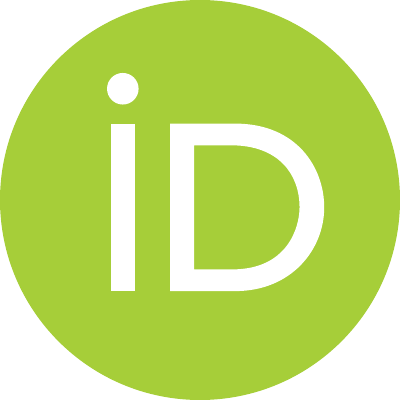}}}}

\usepackage{cite} 
\usepackage{mciteplus}

\usepackage{longtable} 
\usepackage{subcaption}
\begin{document}

\renewcommand{\thefootnote}{\fnsymbol{footnote}}
\setcounter{footnote}{1}


\begin{titlepage}
\pagenumbering{roman}

\vspace*{-1.5cm}
\centerline{\large EUROPEAN ORGANIZATION FOR NUCLEAR RESEARCH (CERN)}
\vspace*{1.5cm}
\noindent
\begin{tabular*}{\linewidth}{lc@{\extracolsep{\fill}}r@{\extracolsep{0pt}}}
\ifthenelse{\boolean{pdflatex}}
{\vspace*{-1.5cm}\mbox{\!\!\!\includegraphics[width=.14\textwidth]{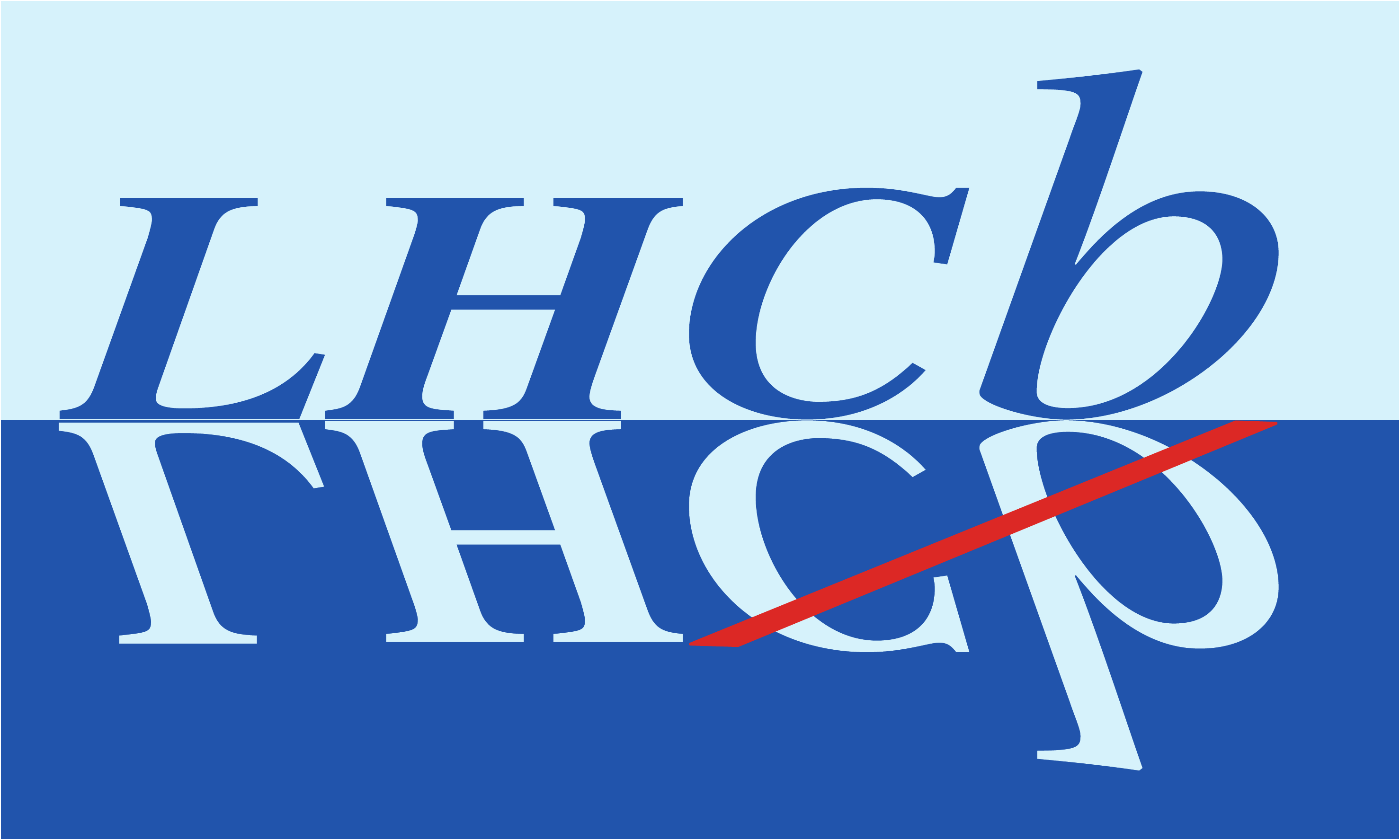}} & &}%
{\vspace*{-1.2cm}\mbox{\!\!\!\includegraphics[width=.12\textwidth]{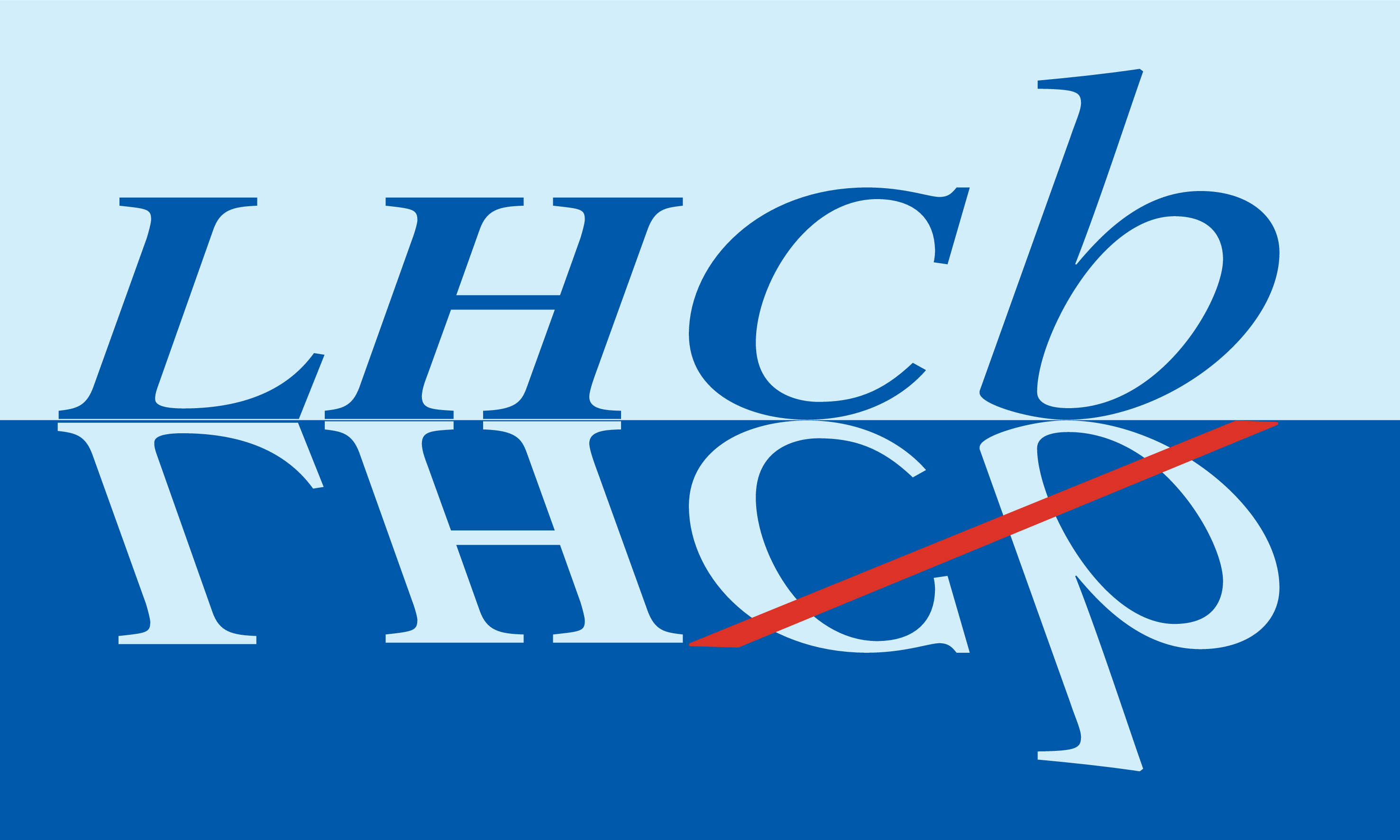}} & &}%
\\
 & & CERN-EP-2023-086 \\  
 & & LHCb-PAPER-2022-053 \\  
 & & 24 May 2023 \\ 
 & & \\
\end{tabular*}

\vspace*{1.1cm}

{\normalfont\bfseries\boldmath\huge
\begin{center}
  \papertitle 
\end{center}
}

\vspace*{0.7cm}

\begin{center}

\paperauthors\footnote{Authors are listed at the end of this paper.}
\end{center}

\vspace{\fill}

\begin{abstract}
  \noindent
    The mass difference between the ${\mathchar"710A}^-_{\ensuremath{b}}$ and ${\mathchar"7104}^-_{\ensuremath{b}}\xspace$ baryons is measured using proton-proton collision data collected by the LHCb experiment, corresponding to an integrated luminosity of $9\ensuremath{\,\text{fb}^{-1}}\xspace$, and is found to be 
    \begin{align*}
        m({\mathchar"710A}^-_{\ensuremath{b}})- m({\mathchar"7104}^-_{\ensuremath{b}}) =  248.54 \pm 0.51\ensuremath{\text{\,(stat)}}\xspace \pm 0.38\ensuremath{\text{\,(syst)}}\xspace  \ensuremath{\text{\,Me\kern -0.1em V\!/}c^2}.
    \end{align*}
    The mass of the ${\mathchar"710A}^-_{\ensuremath{b}}\xspace$ baryon is measured to be
    \begin{align*}
        m({\mathchar"710A}^-_{\ensuremath{b}})= 6045.9 \pm 0.5\ensuremath{\text{\,(stat)}}\xspace \pm 0.6\ensuremath{\text{\,(syst)}}\xspace \ensuremath{\text{\,Me\kern -0.1em V\!/}c^2}.
    \end{align*}
    This is the most precise determination of the ${\mathchar"710A}^-_{\ensuremath{b}}\xspace$ mass to date.
    In addition, the production rate of ${\mathchar"710A}^-_{\ensuremath{b}}\xspace$ baryons relative to that of ${\mathchar"7104}^-_{\ensuremath{b}}\xspace$ baryons is measured for the first time in $\ensuremath{pp}$ collisions, using an LHCb dataset collected at a center-of-mass energy of $13 \ensuremath{\text{\,Te\kern -0.1em V}}\xspace$ and corresponding to an integrated luminosity of $6\ensuremath{\,\text{fb}^{-1}}\xspace$. Reconstructing beauty baryons  in the kinematic region $2 < \eta < 6$ and $p_T < 20\ensuremath{\text{\,Ge\kern -0.1em V\!/}}c$ with 
     their decays  to a  ${\ensuremath{{\ensuremath{J}\mskip -3mu/\mskip -2mu\ensuremath{\psi}}}}\xspace$ meson and a hyperon, 
     the ratio
    \begin{align*}
        \frac{f_{{\mathchar"710A}^-_{\ensuremath{b}}}}{f_{{\mathchar"7104}^-_{\ensuremath{b}}}}\times\frac{\mathcal{B}({\mathchar"710A}^-_{\ensuremath{b}} \to {\ensuremath{{\ensuremath{J}\mskip -3mu/\mskip -2mu\ensuremath{\psi}}}} {\mathchar"710A}^-)}{\mathcal{B}({\mathchar"7104}^-_{\ensuremath{b}} \to {\ensuremath{{\ensuremath{J}\mskip -3mu/\mskip -2mu\ensuremath{\psi}}}} {\mathchar"7104}^-)} =
        0.120 \pm 0.008 \ensuremath{\text{\,(stat)}}\xspace \pm 0.008 \ensuremath{\text{\,(syst)}}\xspace,
    \end{align*}
    is obtained, where $f_{{\mathchar"710A}^-_{\ensuremath{b}}}$ and $f_{{\mathchar"7104}^-_{\ensuremath{b}}}$ are the fragmentation fractions of $\ensuremath{b}$ quarks into ${\mathchar"710A}^-_{\ensuremath{b}}\xspace$ and ${\mathchar"7104}^-_{\ensuremath{b}}\xspace$ baryons, respectively, and $\mathcal{B}$ represents the branching fractions of their respective decays.\\
\end{abstract}

\vspace*{1.cm}

\begin{center}
Published in Phys.Rev.D 108 (2023) 5, 052008
\end{center}

\vspace{\fill}

{\footnotesize 
\centerline{\copyright~\papercopyright. \href{\paperlicenceurl}{\paperlicence}.}}
\vspace*{2mm}

\end{titlepage}


\newpage
\setcounter{page}{2}
\mbox{~}


\renewcommand{\thefootnote}{\arabic{footnote}}
\setcounter{footnote}{0}

\cleardoublepage


\pagestyle{plain} 
\setcounter{page}{1}
\pagenumbering{arabic}


\section{Introduction}
\label{sec:Introduction}

The quark model predicts fifteen \bquark-baryon states, containing one \bquark quark and two lighter quarks (\uquark, \dquark or \squark), that are neither orbitally nor radially excited. Four of these states, namely \Lb, \Xibz, \Xibm and \Omegab, are particularly interesting since they can only decay through the weak interaction, leading to lifetimes of about $1.5\ps$~\cite{PDG2022}. Due to the relatively high masses of the \bquark baryons, currently only high-energy accelerators such as the Large Hadron Collider (LHC) copiously produce all their types. As a consequence, the current experimental knowledge on the \bquark-baryon mass hierarchy, properties and decay modes is limited mainly to the results of hadron-collider experiments. 

The heaviest of the four weakly decaying baryons, the \Omegab state with $ssb$ quark content in the constituent quark model, is the least-well studied. While both the DØ and CDF collaborations claimed the observation of the \OmegabToOmegaJPsi decay, the reported mass values, $6165 \pm 10 \stat \pm 13 \syst \mevcc$ by DØ~\cite{D0:2008sbw} and $6054.4 \pm 6.8 \stat \pm 0.9 \syst \mevcc$ by CDF~\cite{CDF:2009sbo}, differ by more than six standard deviations. The situation was clarified by the LHCb experiment in 2013~\cite{LHCb-PAPER-2012-048} with the confirmation of the state observed by CDF using the same decay mode. In 2014, the CDF experiment observed a second decay mode, $\Omegab\to \Omegac \pim$~\cite{CDF:2014mon}, confirmed by LHCb in 2016~\cite{LHCb-PAPER-2016-008}. 
Both CDF and LHCb have measured the mass and lifetime of the \Omegab baryon. However, in both cases the measurements were limited by the available datasets having yields as low as a few tens of signal candidates. Recently, LHCb published a more precise measurement of the \Omegab baryon mass via its decay mode $\Omegab \to \Xicp \Km \pim$~\cite{LHCb-PAPER-2021-012}.

This paper presents a measurement of the \Omegab baryon mass with the \OmegabToOmegaJPsi decay (charge conjugation is implied throughout this paper), using the full LHCb proton-proton (\pp)~collision dataset collected from 2011 to 2018, corresponding to an integrated luminosity of $9\invfb$, at center-of-mass energies of $7,8$ and $13\tev$. 
Reconstructing the \Omegab baryon through the decay chain \OmegabToOmegaJPsi, $\Omegam\to\Lambdares \Km$ and $\Lambdares\to p \pim$ is particularly advantageous since the cascade topology of the decay of two long-lived hyperons cannot be mimicked by any mesonic decay. This final state leads to very clean signal samples and therefore is well suited for a mass measurement. 

This paper also presents the first measurement of the relative production rates of the \Omegab and \Xibm baryons at the LHC, in $pp$ collisions at a center-of-mass energy of $13\tev$, using a dataset corresponding to an integrated luminosity of $6\invfb$. 
In order to measure the absolute branching fractions of \Omegab decays at the LHC, it is necessary to consider the probability of the \bquark-quark hadronization into the \Omegab baryon, denoted as $f_{\Omegab}$, in  proton-proton collisions. This quantity is also required to measure the total \bquark-quark production cross-section at the LHC, but until now only phenomenological estimates were available~\cite{LHCb-PAPER-2016-031}. 
Experimentally, the quantity accessible at the LHC is the product of the production rate of a baryon and the branching fraction of its decay to a specific final state. This makes it difficult to measure the absolute production rates or absolute branching fractions at hadron colliders without external input. For the \Lb baryon, the production rate was measured at LHCb relative to $B$ mesons. The most precise measurement utilizes the predicted similarity of \bquark-hadron semileptonic decay widths~\cite{LHCb-PAPER-2018-050}. At the moment, extending this technique to the \Omegab baryon is not possible: although semileptonic decays of \Omegab have been used by LHCb to obtain a sample of \Omegac baryons~\cite{LHCb-PAPER-2018-028}, the absolute branching fractions for the \Omegac decays are not known~\cite{PDG2022}. 

The CDF experiment has measured the relative production rates of the \Omegab, \Lb and \Xibm baryons multiplied by their relative decay rates to the final states with a \jpsi meson and a hyperon, in $p\bar{p}$ collisions at the center-of-mass energy of $1.96\tev$ at the \tevatron~\cite{CDF:2009sbo}.\footnote{A similar measurement was also performed by the DØ experiment~\cite{D0:2008sbw}, but since the mass value of the state observed by DØ is incompatible with other experiments, this result is omitted.} 
Since the \tevatron and the LHC use different colliding particles and center-of-mass energies, it is necessary to perform an independent measurement of the relative production fraction at the LHC. The decays \OmegabToOmegaJPsi and \XibToXiJPsi are well suited to perform such a measurement, given their similar topologies and final-state particles. Unfortunately, the relative production fractions cannot be disentangled from the ratio of branching fractions due to the limited knowledge of the absolute branching fractions of the chosen decays or their ratio. Therefore, the relative production rate of the \Omegab and \Xibm baryons is determined with the  ratio
\begin{align}
    R \equiv \frac{f_{\Omegab}}{f_{\Xibm}}\times\frac{\mathcal{B}(\OmegabToOmegaJPsi)}{\mathcal{B}(\XibToXiJPsi)}.
    \label{eqn:ratio}
\end{align}

In the past, LHCb has used a similar approach to measure the relative production rate of the \Xibm and \Lb baryons~\cite{LHCb-PAPER-2018-047} using their decays \XibToXiJPsi and \LbToLJPsi. These two decay modes are linked by an $\mathrm{SU}(3)$ symmetry, which simplifies the theoretical interpretation of this measurement in terms of $f_{\Xibm}/f_{\Lb}$. 
There is no such symmetry relation between the \OmegabToOmegaJPsi and \XibToXiJPsi  decays that can be used to determine $f_{\Omegab}/f_{\Xibm}$, because the \Omegam baryon is not a member of the same baryon multiplet as the \Xim and \Lz baryons. 
Therefore, further theoretical input is needed to directly access the production ratio $f_{\Omegab}/f_{\Xibm}$. Predictions of branching fractions of the \OmegabToOmegaJPsi and \XibToXiJPsi decays have been obtained, among others, with the nonrelativistic quark model~\cite{Cheng:1996cs}, covariant confined quark model~\cite{Gutsche:2018utw}, and light-front quark model~\cite{Hsiao:2021mlp} approaches. The lack of consistency among the different models requires further study. Therefore, the primary quantity measured in this paper is the ratio defined in Eq.~\eqref{eqn:ratio}. %

This paper is organized as follows: after a short introduction of the LHCb detector and its data processing chain in Sec.~\ref{sec:Detector}, the candidate selection and background suppression is described in Sec.~\ref{sec:Selection}. The mass measurement is described in Sec.~\ref{sec:mass} and the relative production rate measurement in Sec.~\ref{sec:production}. This paper concludes with a discussion of the results in Sec.~\ref{sec:Conclusion}.

\section{LHCb detector and datasets}
\label{sec:Detector}
The \lhcb detector~\cite{LHCb-DP-2008-001,LHCb-DP-2014-002} is a single-arm forward spectrometer covering the \mbox{pseudorapidity} range $2<\eta <5$, designed for the study of particles containing \bquark or \cquark quarks. The detector includes a high-precision tracking system consisting of a silicon-strip vertex detector (VELO) surrounding the $pp$ interaction region~\cite{LHCb-DP-2014-001}, a large-area silicon-strip detector located upstream of a dipole magnet with a bending power of about $4{\mathrm{\,Tm}}$, and three stations of silicon-strip detectors and straw drift tubes~\cite{LHCb-DP-2013-003,LHCb-DP-2017-001} placed downstream of the magnet. The tracking system provides a measurement of the momentum, \ptot, of charged particles with a relative uncertainty that varies from 0.5\% at low momentum to 1.0\% at 200\gevc. The minimum distance of a track to a primary $pp$ collision vertex (PV), the impact parameter (IP), is measured for charged particles reconstructed in all tracking subdetectors with a resolution of $(15+29/\pt)\mum$, where \pt is the component of the momentum transverse to the beam, in\,\gevc. Different types of charged hadrons are distinguished using information from two ring-imaging Cherenkov detectors~\cite{LHCb-DP-2012-003}. 
Muons are identified by a system composed of alternating layers of iron and multiwire proportional chambers~\cite{LHCb-DP-2012-002}.

The datasets considered for this measurement were collected by LHCb between 2011-2012 at center-of-mass energies of $7$ and $8\tev$ (Run 1), corresponding to an integrated luminosity of $3 \invfb$, and between 2015-2018 at a center-of-mass energy of $13\tev$ (Run 2), corresponding to an integrated luminosity of $6\invfb$. For the relative production rate measurement, only the Run 2 dataset is used, where the simulated samples describe the reconstruction of long-lived hyperons more precisely. 
The online event selection is performed by a trigger~\cite{LHCb-DP-2012-004}, which consists of a hardware stage, based on information from the calorimeters and the muon system, followed by a software stage, which applies a full event reconstruction. At the hardware stage, the event is required to have a muon with a high transverse momentum or a dimuon pair with a high product of transverse momenta of the two muons. At the software stage, the dimuon pair must have a significant displacement from the PV.

The momentum scale is calibrated using samples of $\decay{\jpsi}{\mumu}$ 
and $\decay{\Bu}{\jpsi\Kp}$~decays collected concurrently
with the~data sample used for this analysis~\cite{LHCb-PAPER-2012-048,LHCb-PAPER-2013-011}.
A relative uncertainty of $3 \times 10^{-4}$ is estimated using samples of other fully reconstructed $\bquark$~hadrons, $\PUpsilon$~mesons and $\KS$~mesons.

Simulated samples of \OmegabToOmegaJPsi and \XibToXiJPsi decays are used to model detector effects and selection requirements. 
In the simulation, $pp$ collisions are generated using \pythia~\cite{Sjostrand:2007gs,*Sjostrand:2006za} with a specific \lhcb configuration~\cite{LHCb-PROC-2010-056}. As the \Omegab production rate in \pythia is very low, its generation is inefficient. Therefore, the \Omegab baryon is simulated as a \Xibm baryon with the mass and lifetime values altered to match the \Omegab world-average values. 
Decays of unstable particles are described by \evtgen~\cite{Lange:2001uf}, in which final-state radiation is generated using \photos~\cite{davidson2015photos}. The decays of the \Omegab and \Xibm baryons are simulated according to available phase space. The decays of the \Omegam, \Xim and \Lz hyperons are simulated  with a dedicated helicity-amplitude model, using the latest world-average values of parity-violating asymmetry parameters in these decays~\cite{PDG2022}. The interaction of the generated particles with the LHCb detector, and its response, are implemented using the \geant toolkit~\cite{Allison:2006ve, *Agostinelli:2002hh} as described in Ref.~\cite{LHCb-PROC-2011-006}.

\section{Candidate selection}
\label{sec:Selection}
The decays \OmegabToOmegaJPsi and \XibToXiJPsi share a particular topology, which is distinct from most other processes occurring in the LHCb detector. The cascade topology of the $\Xim \to \Lz [\to p \pim] \pim$ and $\Omegam \to \Lz [\to p \pim] \Km$ decays cannot be produced by any meson decays and provides a clean signature for hyperon decays at LHCb. This feature is exploited in the design of the selection requirements used in this analysis.

In this paper, reconstructed charged hadrons are classified into two categories. The \textit{long-track} category (L) refers to tracks that have reconstructed segments in both the VELO and the subsequent tracking stations. The \textit{downstream-track} category (D) consists of particles with trajectories that are not reconstructed in the VELO, and only include information from the tracking detectors just before and after the LHCb magnet. While most of the reconstructed charged particles produced in the $pp$ collisions form long tracks, the decay products of long-lived hyperons tend to be reconstructed as downstream tracks. Due to the presence of tracking information from the VELO detector, the trajectories, and hence the IP, of long tracks are measured with better precision than those of downstream tracks. Since the \Omegam and \Xim baryons are long lived and their decay products include another long-lived particle, the \Lz hyperon, the majority of the signal candidates have at least two downstream tracks. 

Signal \Omegab and \Xibm candidates are built by combining an \Omegam or \Xim hyperon candidate, respectively, with a \jpsi meson candidate. The \jpsi candidate is reconstructed from two oppositely charged particles that are reconstructed as long tracks, are identified as muons by the muon system and are assigned a muon mass hypothesis. The dimuon invariant mass is required to be within the range 3000--3150\mevcc.
The \Omegam candidates are reconstructed via the $\Lz K^-$ decay mode, while the \Xim candidates are reconstructed via the $\Lz\pim$ decay mode. The \Lz hyperon candidate is built from a pair of oppositely charged  particles, both forming either long or downstream tracks, that are assigned the proton and pion hypotheses, and originate from a common displaced vertex. The \Lz candidate is subsequently combined with a displaced charged particle, which is reconstructed as either a long or a downstream track, and is assigned a kaon (for the \Omegam) or a pion (for the \Xim) mass hypothesis. As a result, the \Xim and \Omegam hyperons can be reconstructed in three track-type categories: 
 LLL, where the kaon (pion) of the $\Omegares^-$ ($\Xires^-$) decay, and the pion and proton of the \Lambdares decay are all reconstructed from long tracks; LDD, where the \Lambdares decay products are reconstructed from downstream tracks; and DDD, where all the hyperon decay products are reconstructed from downstream tracks. The proportion of the LLL and LDD categories is larger in the \Omegab decay than in the \Xibm decay, due to the  lifetime of the \Omegam hyperon being about half of the \Xim lifetime. 

The \Omegam, \Xim and \Lz candidates are  required to have a decay vertex of good quality and a decay time greater than $2\ps$. The \Lz candidate must have an invariant mass within $8\mevcc$ around the known \Lz baryon mass~\cite{PDG2022}. The \Omegam and \Xim candidates are required to have an invariant mass within $10 \mevcc$ around their known world-average values~\cite{PDG2022}, where the mass of the \Lz candidate is constrained to its world-average value. Together, all these requirements preserve more than 90\% of reconstructed signal candidates in both channels. 
To further suppress the background, the \Omegam and \Xim hyperons are both required to have \pt above $1\gevc$; this requirement preserves about 93\% of signal candidates in each decay mode. The \Omegab and \Xibm candidates are also required to form a good vertex which is significantly displaced from the PV, and their momentum vectors must be consistent with originating from the PV. Since the \Xibm decay includes two pions, it is ensured that the two pion tracks are not duplicates reconstructed from the same particle. This is done by requiring the angle between the two pions in the laboratory frame to be greater than 0.5~mrad. 
This requirement preserves more than 99\% of signal and for consistency, is also applied between the kaon and pion tracks of the \Omegam decay.

For the \XibToXiJPsi decay, the background of \LbToLJPsi decays combined with a random pion is an important contribution. This background can pollute the region below the signal peak, as $m(\Lb)+m(\pim) < m(\Xibm)$. However, for the \Omegab case with ${m(\Lb)+m(\Km) > m(\Omegab)}$ the equivalent background only contributes to the far upper sideband and can therefore be neglected.
To suppress this background in the \Xibm decays, the pion from the \Xim decay is required to have a significant displacement from the PV. The equivalent displacement requirement is applied to the kaon of the \Omegam decay to improve the cancellation of systematic uncertainties. These requirements retain about 93\% of signal candidates in each decay mode. 

A possible source of background for \OmegabToOmegaJPsi decays are \XibToXiJPsi decays, where a pion is misidentified as a kaon. Conversely, in the \XibToXiJPsi case, the background with a kaon misidentified as a pion is negligible due to the low \Omegab production cross-section. A highly efficient way to suppress this background is by applying a strict kaon identification requirement to the kaon candidate. This approach is used for the mass measurement. 
However, a precise calibration of the efficiency for this requirement is difficult to perform when the kaon is reconstructed as a downstream track. 
Therefore, an alternative approach is used for the measurement of the relative production rate: the invariant mass of the \Lz\Km combination is calculated under the pion mass hypothesis for the kaon, and the resulting \Xim mass peak is vetoed. This preserves about 97\% of \Omegab candidates. 

The contribution from partially reconstructed $\Xib \to \jpsi\Xires(1530) $ decays, where the $\Xires(1530)$ resonance decays to a $\Xim \pi$ final state, is expected to be negligible due to non-overlapping wave functions of the \Xib and $\Xires(1530)$ baryons~\cite{Oliver:2010im}. 
Higher $\Xires$ states, located above the $\Lz K$ threshold, have relatively small decay widths to the $\Xim \pi$ channel. These partially reconstructed processes fail the selection requirements of this measurement due to significant missing energy. 
Furthermore, all excited $\Omegares^{-}$ states are located above the $\Xires K$ threshold~\cite{Liu:2019wdr,PDG2022} and have low partial widths for the $\Omegares \gamma$ and isospin-violating $\Omegares \piz$ decays. 
Therefore, no significant partially reconstructed background is expected in the two analyzed datasets.

The final states of interest can also be reached by Cabibbo-suppressed processes such as $\Xibm\to\jpsi\Sigmares(1385)^-$ and $\Omegab\to\jpsi\Xires(1690)^-$, followed by strong decays such as $\Sigmares(1385)^-\to\Lz\pim$ and $\Xires(1690)^-\to\Lz \Km$. 
Since the hadronic resonances produced in these decays are short-lived, these processes are suppressed by the displacement requirement applied on the hyperon candidates. 
Therefore, the dominant remaining background is combinatorial background, consisting mostly of \jpsi mesons combined with random tracks.

To improve the invariant-mass resolution, an additional kinematic fit is performed on each candidate~\cite{Hulsbergen:2005pu}, where the parent hadron is required to be consistent with originating from the primary vertex. Furthermore, the invariant masses of the \jpsi, $\Lz$, \Xim and \Omegam candidates are constrained to their known mass values~\cite{PDG2022}. 
The resulting constrained invariant mass, $m_{\text{constr}}$, has a resolution around $5.4\mevcc$ for the \Omegab decay and $6.0\mevcc$ for the \Xibm decay. The smaller invariant-mass resolution for the \Omegab decay, in comparison to the \Xib decay, is due to the lower energy release in this decay, as well as the lower lifetime of the \Omegam hyperon resulting in a higher proportion of long tracks, which have a better momentum resolution.
Finally, to have a well-defined fiducial region, the \Omegab and \Xibm candidates are required to have $2<\eta<6$ and $\pt <20\gevc$.

\section{Measurement of the \texorpdfstring{$\boldsymbol{\Omegab}$}{Omega_b}-baryon mass}
\label{sec:mass}
The measurement of the \Omegab-baryon mass is performed using an LHCb dataset of $9\invfb$. The collected data are considered separately for the Run 1 and Run 2 samples, which have slightly different invariant-mass resolutions. 
The mass difference $m(\Omegab)-m(\Xibm)$, rather than the absolute \Omegab mass, is extracted from an unbinned maximum-likelihood fit to benefit from canceling systematic uncertainties  due to imprecise calibration of the momentum scale. 
The fit is performed simultaneously to the two run periods and decay channels. It is implemented in the \textsc{RooFit} toolkit~\cite{Verkerke:2003ir} within the \textsc{ROOT} framework~\cite{Brun:1997pa}.

\begin{figure}[b]
    \begin{subfigure}{.5\textwidth}
        \centering
        \includegraphics[height=2.75in]{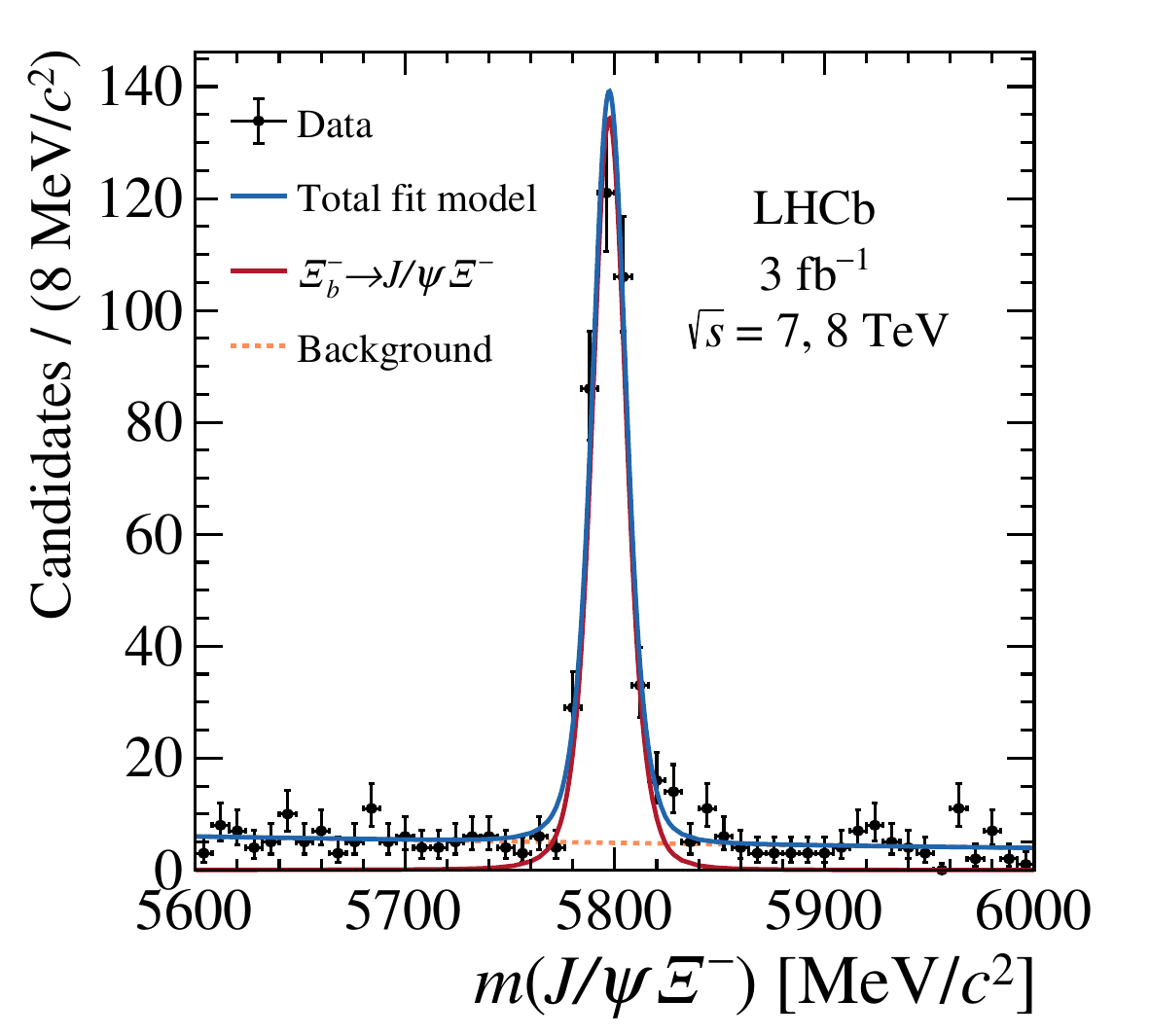}

    \end{subfigure}
    \begin{subfigure}{0.5\textwidth}
        \centering
        \includegraphics[height=2.75in]{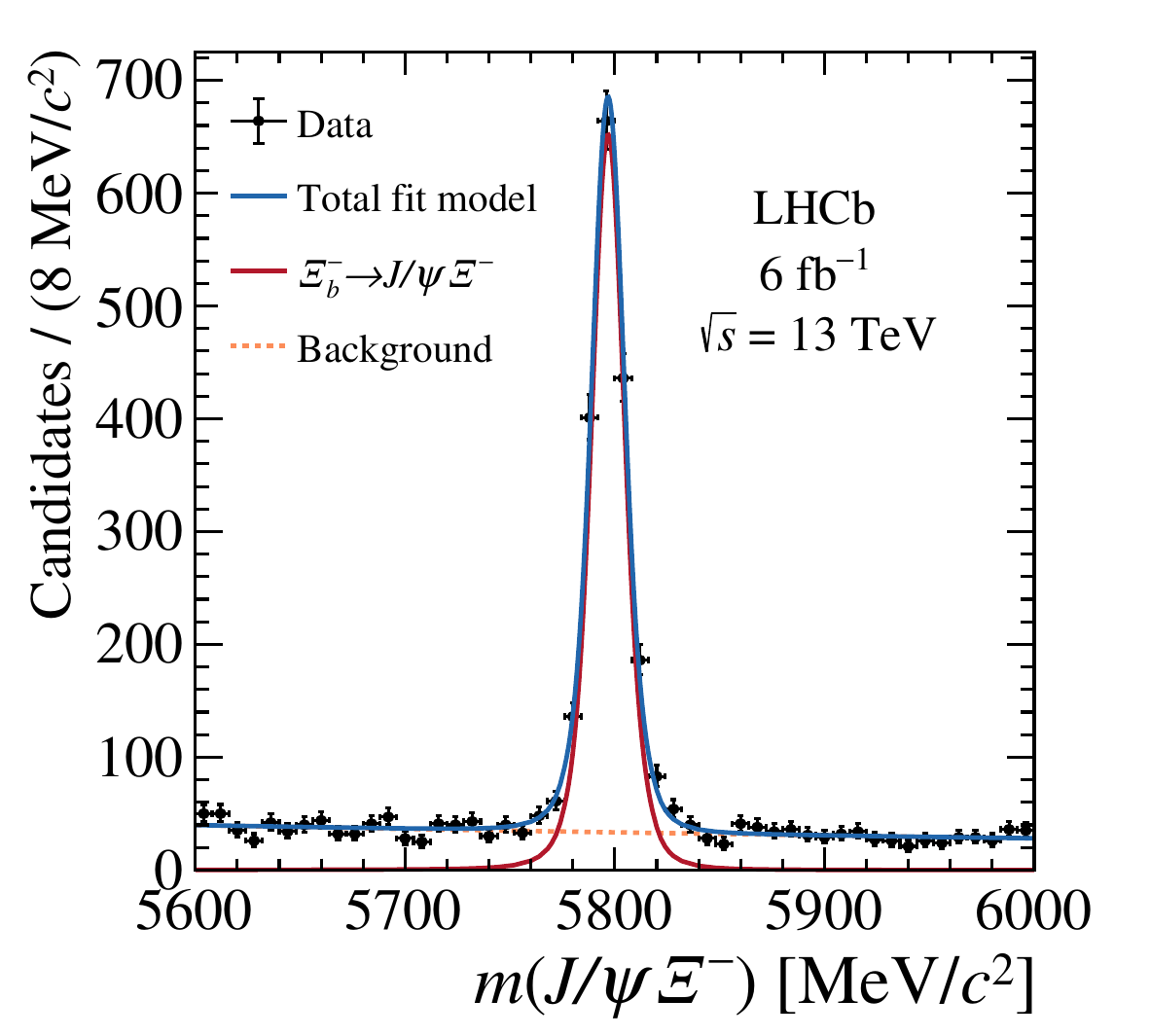}

    \end{subfigure}
    \\
    \begin{subfigure}{0.5\textwidth}
        \centering
        \includegraphics[height=2.75in]{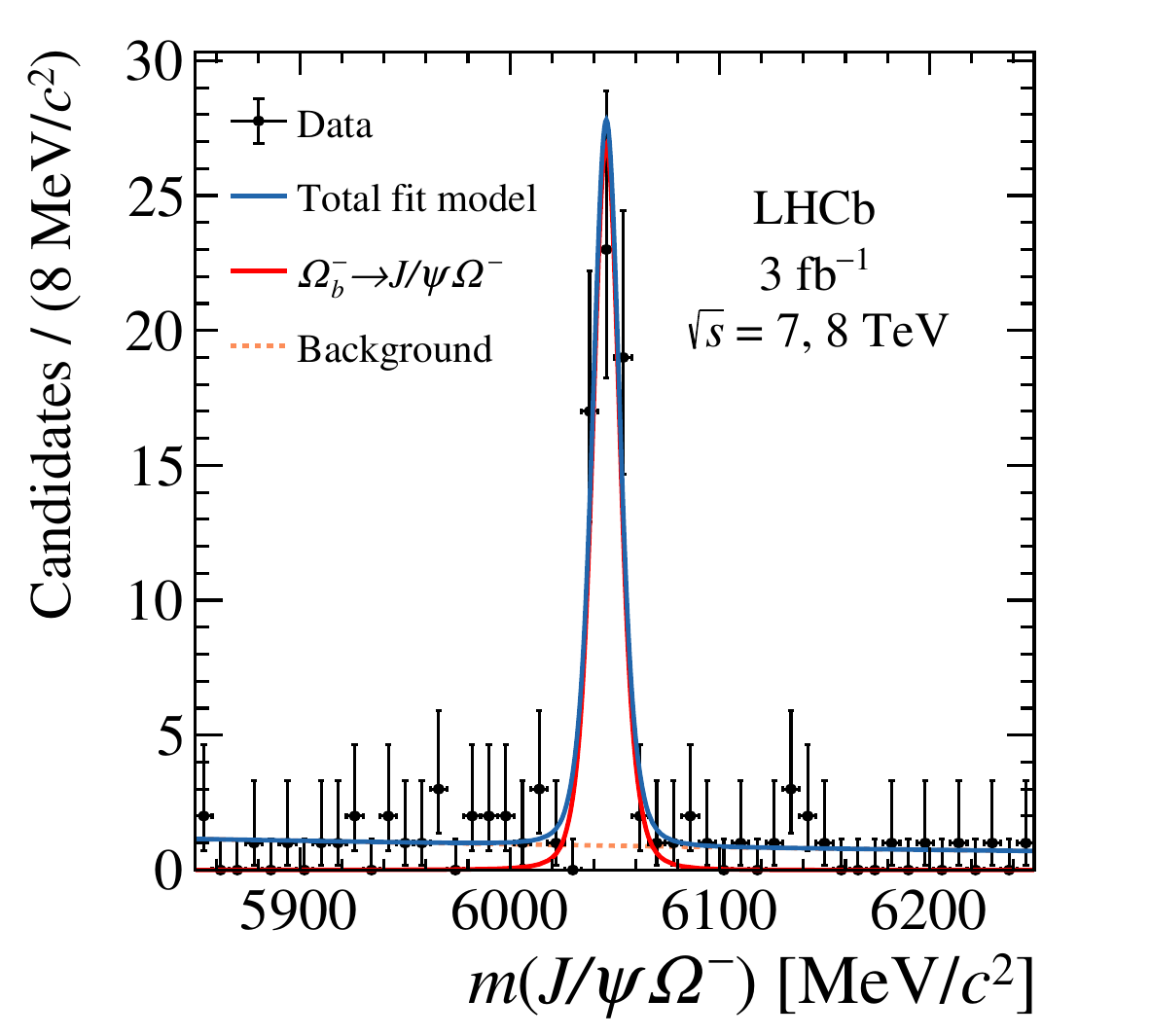}

    \end{subfigure}
    \begin{subfigure}{0.5\textwidth}
        \centering
        \includegraphics[height=2.75in]{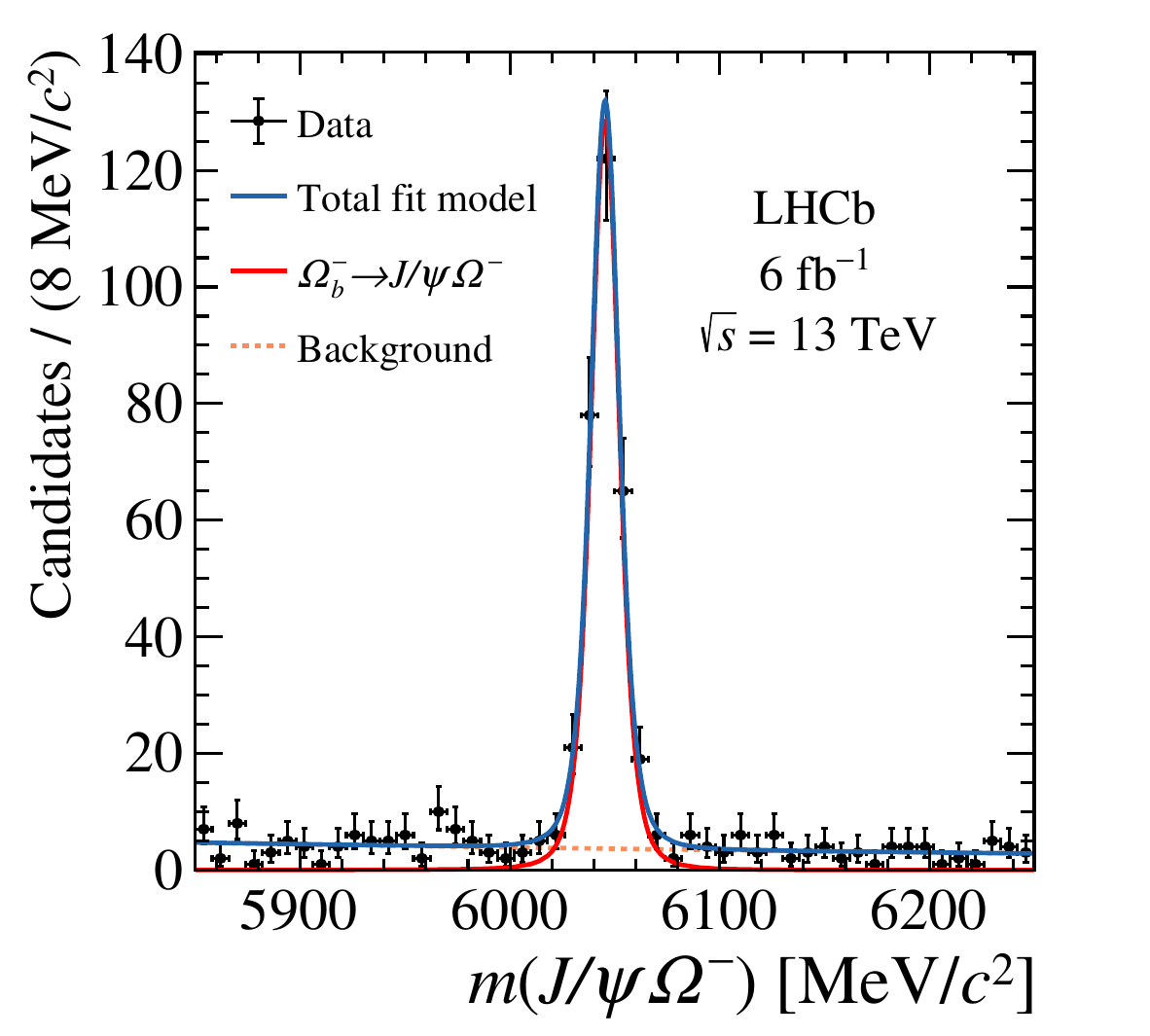}

    \end{subfigure}
    \caption{Invariant-mass distributions in the  (top row) \XibToXiJPsi and (bottom row) \OmegabToOmegaJPsi datasets, in the (left column) Run 1 and (right column) Run 2 data-taking periods. The results of the simultaneous fit used for the mass-difference measurement are overlaid.}
    \label{fig:mfit}
\end{figure}

The resulting invariant-mass spectra are shown in Fig.~\ref{fig:mfit}.
Each signal parametrization is  described by a linear combination of two double-sided Crystal Ball functions~\cite{Skwarnicki:1986xj} with a common mean. The parameters of the signal component are determined from large samples of simulated signal decays; most of them are then fixed in fits to data. The remaining free parameters of the signal in the fit to data are the \Xibm mass for both run periods, the mass difference $m(\Omegab) - m(\Xibm)$, the width of each signal peak, and the signal yields. The background is described in each dataset by an exponential function, with a freely varying normalization and slope. 
Table~\ref{tab:datafit_result_total} shows the resulting signal yields in each dataset.

\begin{table}[t!]
  \centering

  \caption{Obtained values of signal yields from the fit to data.} 
  \label{tab:datafit_result_total}
  \begin{tabular}{c c c} 

        Decay & Dataset & Signal yield \\ 
        \midrule
        \XibToXiJPsi & Run 1 & $377\pm21$ \\ 
        & Run 2 & $1790\pm47\phantom{0}$ \\ 
        \OmegabToOmegaJPsi & Run 1 & $60\pm8$ \\ 
        & Run 2 & $300\pm18$ \\ 

    \end{tabular}
\end{table}

The mass difference from this fit is found to be 
\begin{align*}
    m(\Omegab)-m(\Xibm) = 248.60 \pm 0.51 \stat \mevcc.
\end{align*} 
To exclude dependencies of this result on the different data-taking periods and the track-type categories, the data sample is split into different subsets accordingly, and the mass-difference value measured for each subset. Furthermore, the analysed dataset is split by the charge of the \bquark baryon, as well as by the magnet polarity with which the data was collected. Additional checks are performed by repeating the fit in a narrower invariant-mass range. 
All these cross-checks returned a mass difference in agreement with the baseline value, demonstrating the stability of this result.

The mass difference determined from the fit needs to be corrected for the positive bias ${(0.06\pm0.02)\mevcc}$ observed from the fit to simulation. The uncertainty on this value is due to the size of the simulated sample. The final result therefore is corrected for the bias, and its absolute value is assigned as a systematic uncertainty.
In order to access the final result and determine the value of the \Omegab mass, other sources of systematic uncertainty that do not cancel in the mass difference are studied. They are summarized in Table~\ref{tab:mass_syst}.
\begin{table}[b]
  \centering

  \caption{Systematic uncertainties on the mass-difference measurement. The total systematic uncertainty is obtained by summing all sources in quadrature.} 
  \label{tab:mass_syst}
  \begin{tabular}{l c}

         \multicolumn{1}{c}{Source} & Uncertainty $[\mevcc]$\\
        \midrule
           Momentum scale  & 0.09\\ 
           d$E$/d$x$ correction & 0.01\\
           Hyperon mass  & 0.35\\
           $\LbToLJPsi$ background & 0.10\\
           Fit bias & 0.06\\
           Full fit model & 0.01\\
        \midrule
            Total & 0.38\\

    \end{tabular}
\end{table}

A variation of the momentum scale in simulation within its uncertainty of $\pm0.03\%$ leads to a systematic uncertainty of $0.09\mevcc$ on the mass difference.
The uncertainty on the amount of material assumed in the track reconstruction for the energy loss (d$E$/d$x$) has been found to cancel up to $0.01\mevcc$ in the measurements of mass differences, as long as the number of final-state tracks is the same between the two decay modes~\cite{LHCb-PAPER-2012-048}. 
To propagate the uncertainty on the world-average values of the $\Omegam$ and $\Xim$ masses, the resulting bias on $m_{\text{constr}}$ is estimated. This is performed by varying in simulated datasets the central values assumed in the mass constraints, according to their uncertainties. The average bias on the measured \bquark-baryon mass is evaluated independently for the \Omegab and \Xibm cases, by varying the value of the \Omegam and \Xim mass, respectively; the two uncertainties are then summed in quadrature. It is expected that the small uncertainties on the \Lz and \jpsi mass values cancel in the mass difference. 
The resulting systematic uncertainty of $0.35\mevcc$ is dominated by the uncertainty of the \Omegam mass with $m_{\text{PDG}}(\Omegam)=1672.45\pm 0.29\mevcc$~\cite{PDG2022}. 

The systematic uncertainty associated to possible residual backgrounds from the \mbox{\LbToLJPsi} decay, combined with a random pion, is studied with four alternative approaches. These include three methods to suppress this background: applying a dedicated mass veto on the $\jpsi \Lz$ invariant mass around the known \Lb mass value; requiring a large flight distance of the \Xim candidate; or rejection of good-quality $\jpsi\Lz$ vertices. Alternatively, a dedicated component is included to account for this background in the fit. Averaging the resulting shifts from the baseline result leads to a systematic uncertainty of $0.10\mevcc$ on the mass difference.

While for the signal model a fit bias of $0.06\mevcc$ is observed, no significant additional bias is found for the full fit model. 
The systematic uncertainty related to the chosen fit model is assessed with pseudoexperiments for an alternative background model, where a linear parametrization is assumed instead of the exponential function, or an alternative signal model consisting of a single Crystal Ball distribution. The resulting combined uncertainty due to the fit model is $0.01\mevcc$.

Correcting the mass difference for the value of the bias observed in the fit to simulation, and taking into account the systematic uncertainties, the mass difference is determined to be 
\begin{align*}
    m(\Omegab)-m(\Xibm) = 248.54 \pm 0.51 \stat \pm 0.38 \syst \mevcc.
\end{align*}
The result is in agreement with the previous measurement by LHCb~\cite{LHCb-PAPER-2016-008}, but significantly more precise.

To obtain the value of $m(\Omegab)$, the most precise measurement of the \Xibm baryon mass to date is used: $m(\Xibm) = 5797.33 \pm 0.24 \pm 0.29 \mevcc$~\cite{LHCb-PAPER-2020-032}. This measurement uses $\Xibm\to\Xicz \pim$ decays, and therefore is statistically independent of the \XibToXiJPsi dataset used in this paper. The systematic uncertainties are considered to be fully uncorrelated, except for the uncertainties due to momentum scale calibration and the energy loss correction, which are taken as fully correlated. This leads to a value of the \Omegab baryon mass of
\begin{align*}
   m(\Omegab) = 6045.9 \pm 0.5 \stat \pm 0.6 \syst \mevcc.
\end{align*}
This value is the most precise determination of the \Omegab mass and is consistent with the current world average of $6045.2\pm1.2\mevcc$~\cite{PDG2022}. 
This result supersedes the value reported in Ref.~\cite{LHCb-PAPER-2012-048}.

Combining our value of $m(\Omegab)-m(\Xibm)$ with that in Ref.~\cite{LHCb-PAPER-2016-008}, and assuming that the systematic uncertainties due to momentum scale calibration and the energy loss correction are fully correlated between the two measurements, while all other uncertainties are fully uncorrelated, one obtains
\begin{align}
    [m(\Omegab)-m(\Xibm)]_{\textrm{LHCb comb}} = 248.50 \pm 0.51 \stat \pm 0.37 \syst \mevcc.
\label{eq:average_diff}
\end{align}

In order to obtain the combination of LHCb determinations of the \Omegab baryon mass, the measurement in Ref.~\cite{LHCb-PAPER-2021-012}, which does not use the \Xibm mass as a reference, is combined with the two measurements that rely on the \Xibm mass (Ref.~\cite{LHCb-PAPER-2016-008} and the measurement presented here), which are already combined in Eq.~\ref{eq:average_diff}. This is achieved by first extracting $m(\Omegab)$ from Eq.~\eqref{eq:average_diff} using the \Xibm mass value from Ref.~\cite{LHCb-PAPER-2020-032}. The obtained value of $m(\Omegab)$ is then combined with that from Ref.~\cite{LHCb-PAPER-2021-012}.

The uncertainties due to the momentum-scale and the energy-loss corrections are treated as fully correlated, and all others as uncorrelated. 
This leads to an LHCb combination of the \Omegab baryon mass of
\begin{align*}
    m(\Omegab)_{\textrm{LHCb comb}} = 6045.7\pm0.5 \stat\pm0.6 \syst \mevcc.
\end{align*}

\begin{figure}[t!]
        \centering
        \includegraphics[height=3.in]{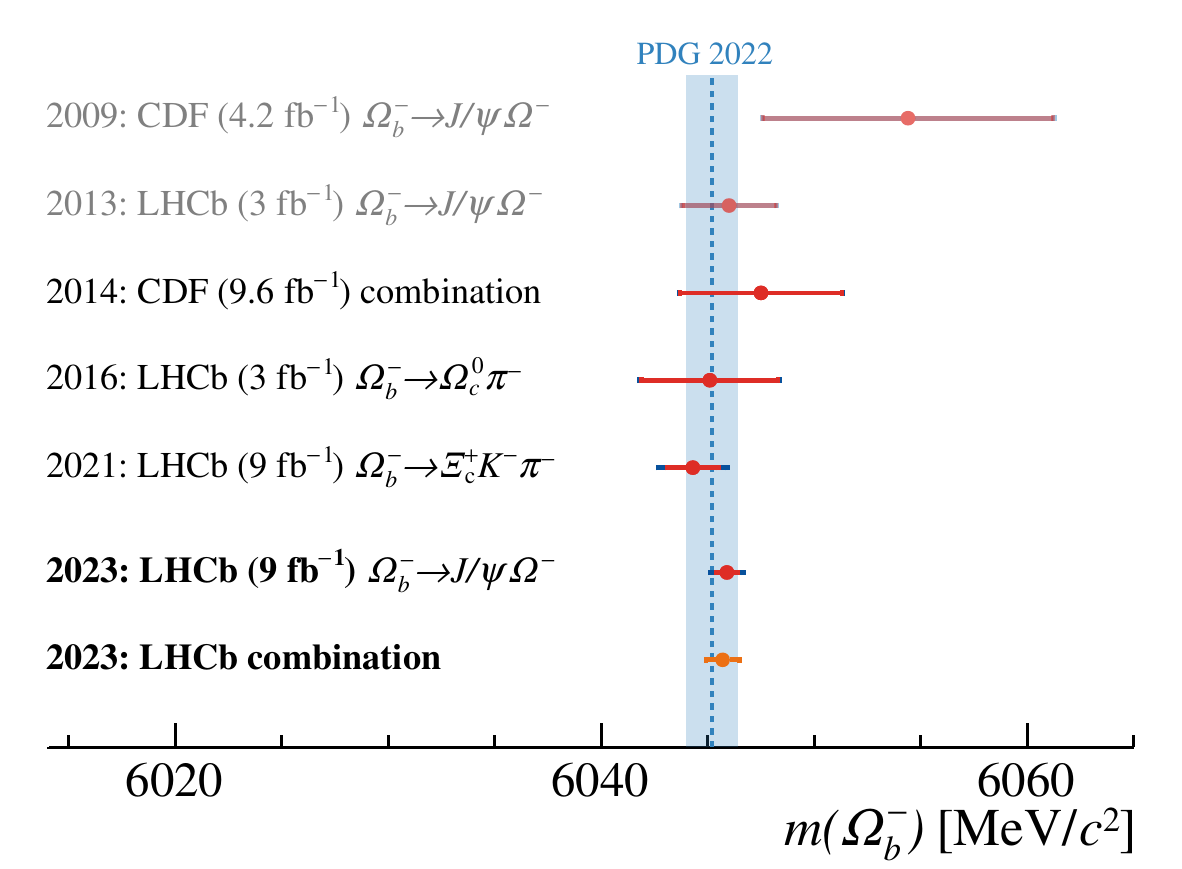}

    \caption{Overview of the \Omegab mass measurements to date (red points)~\cite{CDF:2009sbo,LHCb-PAPER-2012-048,CDF:2014mon,LHCb-PAPER-2016-008,LHCb-PAPER-2021-012}, with corresponding statistical and systematic uncertainties marked in red and blue, respectively, as well as the current world average by the PDG~\cite{PDG2022} (light-blue band). The superseded measurements are shown with lighter colors. The  measurement by the DØ experiment~\cite{D0:2008sbw}  of $6165 \pm 22\mevcc$ is not shown, because it is inconsistent with all others and falls outside the range of this plot. The new LHCb combination is presented in orange, where the total uncertainty is shown.}
    \label{fig:masssummary}
\end{figure}

A summary of \Omegab mass determinations from the CDF and LHCb experiments is presented in Fig.~\ref{fig:masssummary}. 

\section{Relative production of \texorpdfstring{$\boldsymbol{\Omegab}$}{Omega_b} and \texorpdfstring{$\boldsymbol{\Xibm}$}{Xi_b} baryons}
\label{sec:production}

In the Run 1 dataset, the signal yield of the \OmegabToOmegaJPsi decay is limited, and the downstream tracking efficiency is poorly known. Therefore, the relative production ratio of \Omegab and \Xibm baryons is only reported at a center-of-mass energy of $13\tev$, using the Run 2 dataset. 
The relative production ratio of the \Omegab and \Xibm baryons, defined in Eq.~\eqref{eqn:ratio}, is experimentally measured using the observed yields ($N$) of the considered decays, corrected for efficiency ($\epsilon$) and branching fractions ($\mathcal{B}$) of hyperon decays: 
\begin{align}
    R = \frac{N(\OmegabToOmegaJPsi)}{N(\XibToXiJPsi)}\times\frac{    \epsilon(\XibToXiJPsi)}{\epsilon(\OmegabToOmegaJPsi)}\times  \frac{\mathcal{B}(\Xim\to\Lz\pim)}{\mathcal{B}(\Omegam\to\Lz\Km)}.
    \label{eqn:ratioexp}
\end{align}
Since the decays $\Lz\to p\pim$ and $\jpsi\to\mumu$ enter both \bquark-baryon decay chains, their branching fractions cancel in the ratio. The world-average values of ${\mathcal{B}(\Omegam\to\Lz\Km)} = (67.8\pm0.7) \%$ and  ${\mathcal{B}(\Xim\to\Lz\pim)=(99.887\pm0.035)\%}$ are used~\cite{PDG2022}. 
The efficiencies are estimated using the relevant simulation samples. In the ratio, many sources of inefficiency and associated systematic uncertainties cancel due to the very similar kinematics and topologies of the two decay modes. However, this cancellation is not expected for the detector acceptance and reconstruction efficiencies due to the large difference in lifetimes between the \Omegam and \Xim hyperons. 

\begin{figure}[!b]
    \begin{subfigure}{0.5\textwidth}
        \centering
        \includegraphics[height=3in]{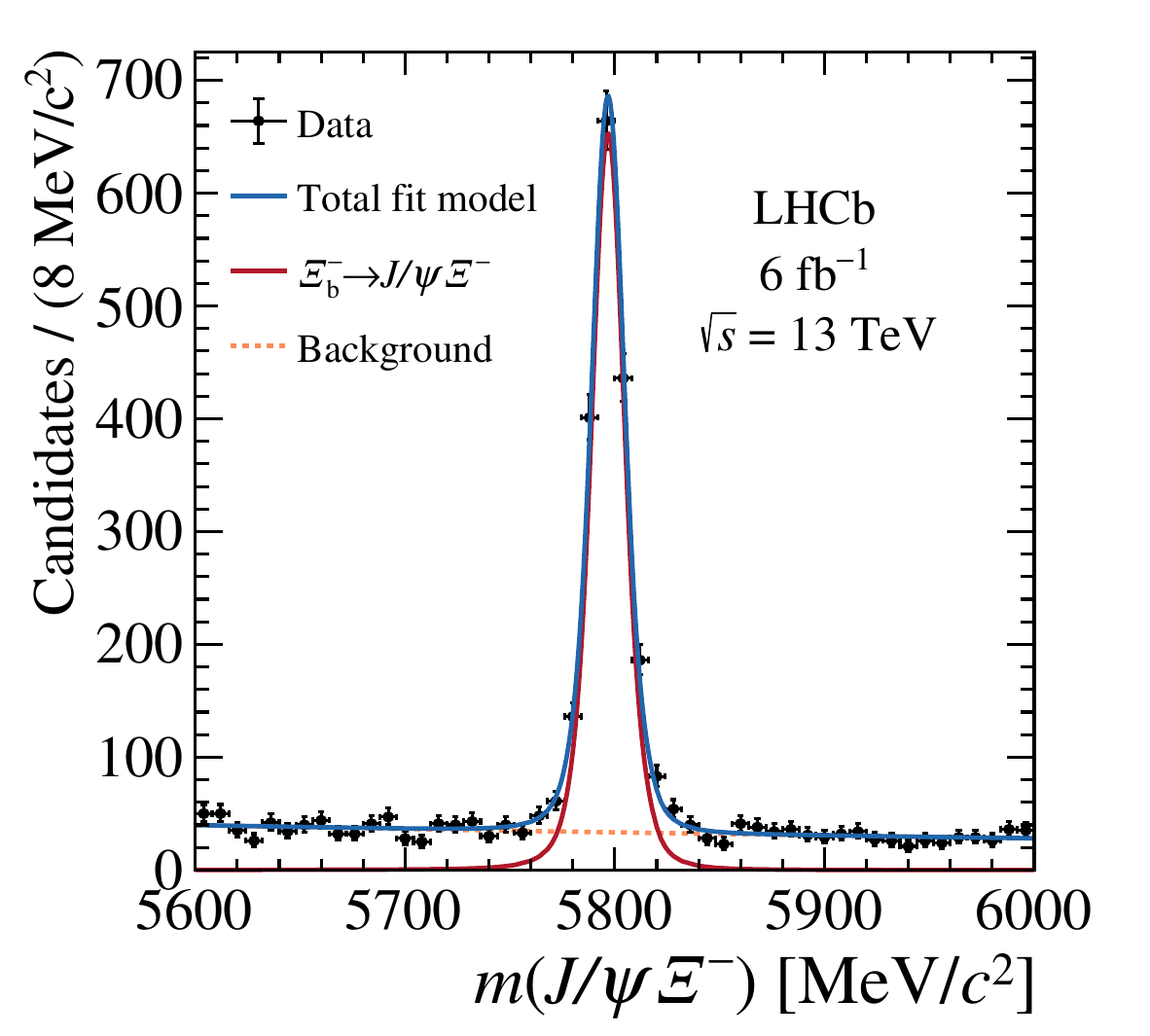}
    \end{subfigure}
    \begin{subfigure}{0.5\textwidth}
        \centering
        \includegraphics[height=3in]{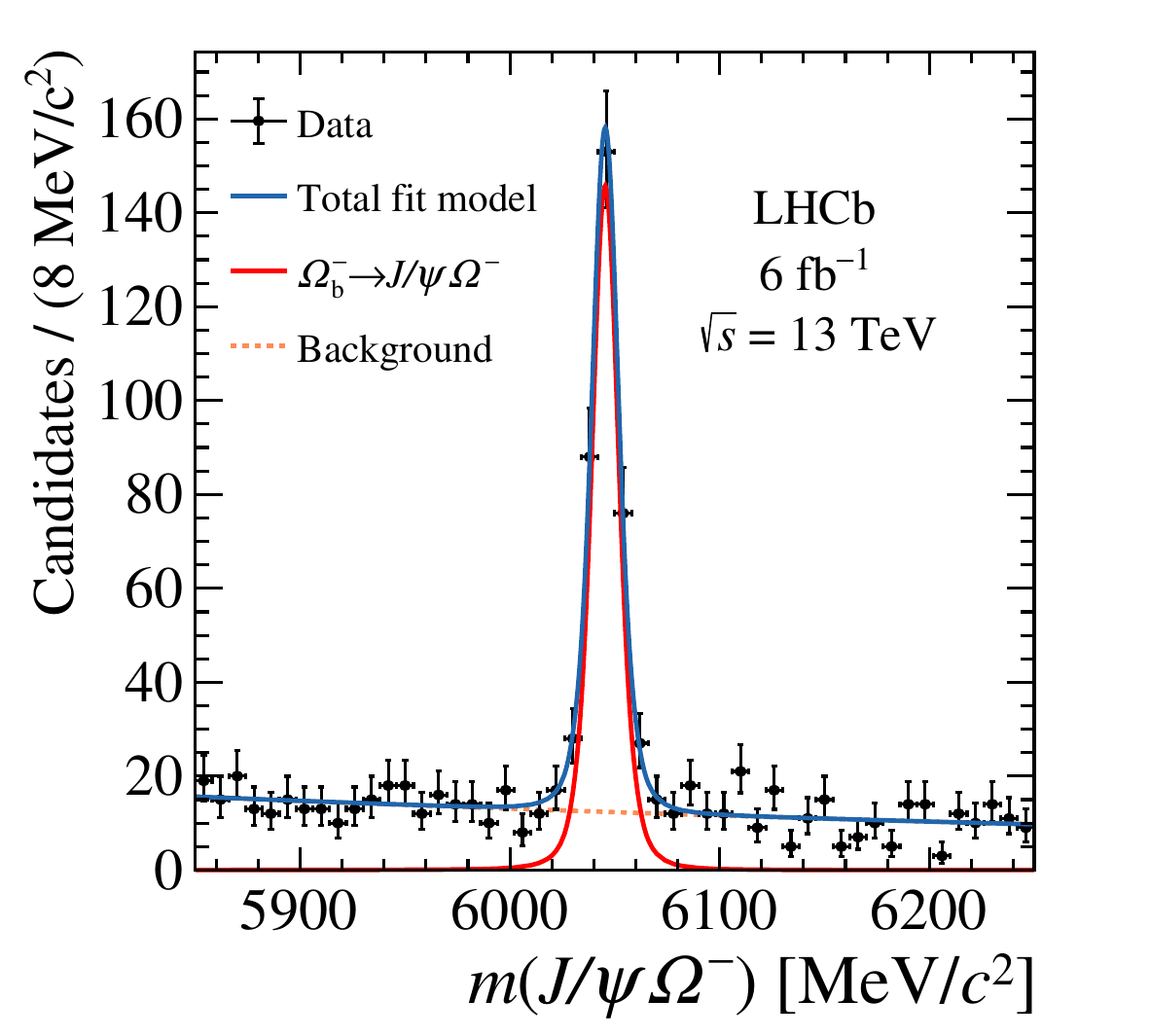}
    \end{subfigure}
    \caption{Invariant-mass distributions in the (left) \XibToXiJPsi and (right) \OmegabToOmegaJPsi \mbox{Run 2} datasets with the results of the simultaneous fit overlaid used for the relative production estimation.
In the \OmegabToOmegaJPsi dataset a \Xim mass veto is applied rather than the kaon identification requirement. 
    }
    \label{fig:prod}
\end{figure}

A simultaneous fit to the Run 2 datasets for both decay modes is performed. The fit setup is nearly identical to that used for the mass measurement, but the signal yield of the \OmegabToOmegaJPsi decays is expressed as a product of the relative production ratio $R$, the yield of \XibToXiJPsi decays and the ratio of efficiencies for the two decay modes. The mass projections of the fit are shown in Fig.~\ref{fig:prod}. 

To determine the ratio in Eq.~\eqref{eqn:ratioexp}, the ratio of efficiencies is evaluated from calibrated simulation samples of the two signal decays. The calibration procedure is performed to correct for known sources of mismodeling in simulation. 
The simulated kinematics of \bquark baryons are calibrated by weighting the distribution of the transverse momenta of the \Omegab and \Xibm baryons to match their distributions in data, after the background contribution is subtracted using the \splot method~\cite{Pivk:2004ty}. The momentum asymmetries, $(p_1 - p_2)/(p_1 + p_2)$, of the decay products of the \Xim and  \Lambdares hyperons are calibrated with a similar approach. For the \Omegam-hyperon decay, a reasonable agreement 
is observed between simulation and data, and therefore no weighting is applied.
Such a calibration of the momentum asymmetry can account for the simplified modeling of the angular distributions in the  simulation of the \bquark-baryon decays according to available phase space. The hardware trigger response is calibrated based on a tag-and-probe approach using $\Bp \to \jpsi \Kp$ decays~\cite{LHCb-PUB-2014-039}. The muon identification response is calibrated with dedicated tag-and-probe $\jpsi \to \mumu$ samples~\cite{LHCb-DP-2018-001}. 
After calibration, good agreement between the simulation and background-subtracted data is observed in variables that describe the decay kinematics and topology.

The calibrated simulation is used to calculate the signal efficiencies. These efficiencies account for the detector acceptance, the detection, reconstruction, trigger and muon identification efficiency as well as the final kinematic and geometric selection.
The resulting efficiencies and their ratio are summarized in Table~\ref{tab:eff}. These efficiency values are found to be higher for the \Omegab decay channel, which is mainly attributed to the shorter lifetime of the \Omegam hyperon as compared to the \Xim hyperon. This results in a larger fraction of the \Omegab decays occurring within the acceptance of the LHCb tracking system. 

\begin{table}[bt]
  \centering

  \caption{Efficiencies obtained from the simulated samples of \OmegabToOmegaJPsi and \XibToXiJPsi decays in Run 2, used to extract the ratio $R$. The uncertainties shown in this table only account for the size of the simulated samples.} 
  \label{tab:eff}
    \begin{tabular}{cc}

    Channel & Efficiency   \\
\midrule
    \XibToXiJPsi &  $(0.923 \pm 0.002) \times 10^{-3}$  \\

    \OmegabToOmegaJPsi & $(2.065 \pm 0.004)\times 10^{-3}$   \\
\midrule
    ratio $\Omegab/\Xibm$ & $2.236 \pm 0.007$    \\

 \end{tabular}
\end{table}

Using the efficiency ratio from Table~\ref{tab:eff}, 
the relative production ratio is determined to be $ R  =0.120\pm0.008$,
where the uncertainty is statistical only. The stability of this result is tested with a dataset-splitting procedure similar to that used in the mass measurement. The obtained results for all subsets are in excellent agreement with the baseline result. As an additional test, the ratio is also extracted with the invariant-mass range in the fit reduced by $50$ or $100\mevcc$ on both sides, yielding a result consistent with the default value.

Possible sources of systematic uncertainty are studied in detail and summarized in Table~\ref{tab:syst_rpsi}. 
\begin{table}[!b]
  \centering
  \caption{Relative systematic uncertainties on $R$. }
  \label{tab:syst_rpsi}
  \begin{tabular}{l c}

  \multicolumn{1}{c}{Source} & Uncertainty [\%] \\
\midrule
  Size of simulated samples &  $0.3$ \\ 
  Calibration of simulation &  $5.5$ \\ 
  Selection criteria &  $0.1$ \\ 
  Lifetimes of \bquark baryons &  $3.1$ \\ 
  Material interactions &  $0.7$ \\ 
  Fit model & $0.8$ \\ 
  External input ($\mathcal{B}$) &  $1.0$ \\ 
      \midrule 
 Total & $6.5$  \\  

 \end{tabular}
\end{table}
The size of the simulated samples, used to estimate the signal efficiencies, leads to a $0.3\%$ relative uncertainty on the efficiency ratio. 
However, the amount of background-subtracted data used to determine the simulation-calibration weights is rather small, especially for the \Omegab decay. The systematic uncertainty associated to this limited amount of data is propagated using a bootstrapping technique~\cite{efron:1979}. All the data candidates used for the determination of the calibration weights are assigned a Poisson-distributed weight of unity mean. 
This entire calibration procedure is repeated 100 times, resulting in 100 alternative efficiency-ratio values. Their RMS leads to a relative systematic uncertainty of $3.9\%$. 
Furthermore, alternative binning schemes are tested for the calibration weights on the simulation. For instance, the kinematics of the \Xibm baryon is alternatively calibrated in two dimensions, $\pt(\Xibm):\eta(\Xibm)$, rather than the default calibration of $\pt(\Xibm)$ alone. For the \Omegab baryon, the available amount of data is too small to perform the same two-dimensional calibration. 
In addition, a correction for the residual data-simulation discrepancies, present after the application of the default calibration chain, was applied, and the associated shift in the final result is treated as another source of systematic uncertainty. A further source of uncertainty originates from imprecise modeling of the downstream tracking efficiency in simulation. To exclude a momentum-dependent bias, the result is tested as a function of the \Xim (\Omegam) momentum, and the observed variation (2.8\%) is assigned as a systematic uncertainty. 
A dedicated systematic uncertainty is also assigned for the calibration procedure of the muon identification efficiency in simulation (0.6\%)~\cite{LHCb-PUB-2016-021}. Overall, the systematic uncertainty related to the calibration of the simulation is found to be $5.5\%$.

The efficiency of the \Xim veto can be incorrectly estimated in simulation if the invariant-mass resolution is not well modeled. This is tested by broadening and narrowing the vetoed mass region and probing the shift in the final result. The associated relative uncertainty is estimated to be $0.1\%$. 

A further uncertainty arises from propagating the uncertainties on the known values of the \Omegab and \Xibm baryon lifetimes~\cite{PDG2022}, and is determined by weighting the decay-time distributions in the simulation samples with a dedicated exponential factor. This results in a $3.1\%$ relative uncertainty on the production ratio, due to the relatively poor knowledge of the \Omegab baryon lifetime. 

Long-lived hyperons can interact with the detector material prior to their decay. 
The rate of these interactions is not precisely modeled in simulation, due to poor knowledge of the cross-sections for hyperons. A dedicated systematic uncertainty is assigned by comparing the hyperon material-interaction probability estimated with two different versions of the \geant simulation toolkit\footnote{The \geant versions used are \texttt{v9r6p4} and \texttt{v10r6p2}.}, which use different values for hadronic interaction cross-sections, amounting to a relative uncertainty of $0.7\%$. 

Several sources of systematic uncertainty related to the fit model are considered. Using pseudoexperiments, the uncertainties related to fit bias and the chosen signal and background models are studied. The final relative systematic uncertainty from the fit model is $0.8\%$. 
Lastly, the uncertainties on the branching fractions of the $\Omegam \to \Lz \Km$ and $\Xim \to \Lz \pim$ decays are propagated to the final result, leading to a relative uncertainty of $1.0\%$.

Accounting for the fit result and the systematic uncertainties, the ratio $R$ is measured to be: 
\begin{align*}
    R = 0.120 \pm 0.008 \stat \pm 0.008 \syst,
\end{align*}
in the kinematic region $2<\eta<6$ and $\pt < 20$ \gevc for the \Xibm and \Omegab baryons. 
Although the ratio of production cross-sections between the \Omegab and \Xibm is not expected to match between the two production environments, a comparison can be useful.
This result is lower than, but consistent with the value measured by the CDF experiment in $p\bar{p}$ collisions at the center-of-mass energy of $1.96\tev$, \mbox{$R = 0.27 \pm 0.12 \stat \pm 0.01 \syst$~\cite{CDF:2009sbo}}. This comparison is limited by the very low signal yield of \OmegabToOmegaJPsi decays in the CDF dataset.

The CDF measurement was reinterpreted, using the predicted value of the branching fraction of the \Omegab decay, to extract an \Omegab production fraction of $f_{\Omegab}\sim 0.5\times 10^{-2}$ at the \tevatron~\cite{Hsiao:2021mlp}. 
Similarly, one can estimate the value of $f_\Omegab / f_\Xibm$ at LHCb using the predictions for the relevant branching fractions.
For instance, Ref.~\cite{Hsiao:2021mlp} predicts $\mathcal{B}(\OmegabToOmegaJPsi) = (5.3 ^{+5.0}_{-3.4}) \times 10^{-4}$ and quotes $\mathcal{B}(\XibToXiJPsi) = (5.1 \pm 3.2) \times 10^{-4}$. 
As the ratio of these two branching fractions is close to unity, the resulting value of $f_\Omegab / f_\Xibm$ is about 0.12. 
However, it should be noted that other predictions of the two branching fractions~\cite{Cheng:1996cs,Gutsche:2018utw,Hsiao:2021mlp} would yield different results.

\section{Conclusion}
\label{sec:Conclusion}
The mass difference between the \Omegab and \Xibm baryons
is measured to be
\begin{align*}
    m(\Omegab)-m(\Xibm) &= 248.54 \pm 0.51 \stat \pm 0.38 \syst \mevcc~%
\end{align*}
using an LHCb dataset corresponding to an intergrated luminosity of $9\invfb$. 
This result is consistent with the previous direct determination of this quantity in Ref.~\cite{LHCb-PAPER-2016-008}, \mbox{$m(\Omegab)-m(\Xibm) = 247.4 \pm 3.2 \pm 0.5\mevcc$}, which uses LHCb data of \bquark-baryon decays to open-charm final states collected at the center-of-mass energies of 7 and 8~\tev. 
The combination of the two results is 
\begin{align*}
    [m(\Omegab)-m(\Xibm)]_{\textrm{LHCb comb}} = 248.50 \pm 0.51 \stat \pm 0.37 \syst \mevcc.
\end{align*}
The value of the \Omegab baryon mass is measured to be
\begin{align*}
m(\Omegab) &= 6045.9 \pm 0.5 \stat \pm 0.6 \syst \mevcc,
\end{align*}
where the value of $m(\Xibm) = 5797.33 \pm 0.24 \pm 0.29 \mevcc$ from Ref.~\cite{LHCb-PAPER-2020-032} is used as a reference. 
This measurement represents the most precise determination of the \mbox{\Omegab-baryon} mass, and is consistent with the current world average of $6045.2\pm1.2 \mevcc$~\cite{PDG2022}. It supersedes the value measured in Ref.~\cite{LHCb-PAPER-2012-048}. \\
The combination of the $m(\Omegab)$ value determined in this paper with the previous LHCb measurements in Refs.~\cite{LHCb-PAPER-2016-008, LHCb-PAPER-2021-012} leads to 
\begin{align*}
    m(\Omegab)_{\textrm{LHCb comb}} = 6045.7\pm0.5 \stat\pm0.6 \syst \mevcc. 
\end{align*}
In addition, the relative production rate of the \Omegab to \Xibm baryons multiplied by the branching fractions of their decays to $\jpsi \Omegam$ and $\jpsi \Xim$ final states is reported for the first time at the LHC. In $pp$ collisions at a center-of-mass energy of $13\tev$, and for \bquark baryons with $2<\eta<6$ and $\pt < 20$ \gevc, this ratio is determined to be 
\begin{align*}
    R = 0.120 \pm 0.008 \stat \pm 0.008 \syst.
\end{align*}
This measurement is lower than, but still in agreement with the value measured by the CDF experiment in $p\bar{p}$ collisions at a center-of-mass energy of $1.96\tev$~\cite{CDF:2009sbo}. 
This new measurement allows for a more precise determination of the total \bquark-quark production cross-section at the LHC. 
Further theoretical studies of the branching fractions of the decays \OmegabToOmegaJPsi and \XibToXiJPsi are needed in order to interpret this result in terms of $f_\Omegab / f_\Xibm$.

\section*{Acknowledgements}

\noindent We express our gratitude to our colleagues in the CERN accelerator departments for the excellent performance of the LHC. We thank the technical and administrative staff at the LHCb
institutes. We acknowledge support from CERN and from the national agencies:
CAPES, CNPq, FAPERJ and FINEP (Brazil); 
MOST and NSFC (China); 
CNRS/IN2P3 (France); 
BMBF, DFG and MPG (Germany); 
INFN (Italy); 
NWO (Netherlands); 
MNiSW and NCN (Poland); 
MEN/IFA (Romania); 
MICINN (Spain); 
SNSF and SER (Switzerland); 
NASU (Ukraine); 
STFC (United Kingdom); 
DOE NP and NSF (USA).
We acknowledge the computing resources that are provided by CERN, IN2P3
(France), KIT and DESY (Germany), INFN (Italy), SURF (Netherlands),
PIC (Spain), GridPP (United Kingdom), 
CSCS (Switzerland), IFIN-HH (Romania), CBPF (Brazil),
Polish WLCG  (Poland) and NERSC (USA).
We are indebted to the communities behind the multiple open-source
software packages on which we depend.
Individual groups or members have received support from
ARC and ARDC (Australia);
Minciencias (Colombia);
AvH Foundation (Germany);
EPLANET, Marie Sk\l{}odowska-Curie Actions and ERC (European Union);
A*MIDEX, ANR, IPhU and Labex P2IO, and R\'{e}gion Auvergne-Rh\^{o}ne-Alpes (France);
Key Research Program of Frontier Sciences of CAS, CAS PIFI, CAS CCEPP, 
Fundamental Research Funds for the Central Universities, 
and Sci. \& Tech. Program of Guangzhou (China);
GVA, XuntaGal, GENCAT and Prog.~Atracci\'on Talento, CM (Spain);
SRC (Sweden);
the Leverhulme Trust, the Royal Society
 and UKRI (United Kingdom).




\addcontentsline{toc}{section}{References}
\bibliographystyle{LHCb}
\bibliography{main,standard,LHCb-PAPER,LHCb-CONF,LHCb-DP,LHCb-TDR}

\newpage
\centerline
{\large\bf LHCb collaboration}
\begin
{flushleft}
\small
R.~Aaij$^{32}$\lhcborcid{0000-0003-0533-1952},
A.S.W.~Abdelmotteleb$^{51}$\lhcborcid{0000-0001-7905-0542},
C.~Abellan~Beteta$^{45}$,
F.~Abudin{\'e}n$^{51}$\lhcborcid{0000-0002-6737-3528},
T.~Ackernley$^{55}$\lhcborcid{0000-0002-5951-3498},
B.~Adeva$^{41}$\lhcborcid{0000-0001-9756-3712},
M.~Adinolfi$^{49}$\lhcborcid{0000-0002-1326-1264},
P.~Adlarson$^{77}$\lhcborcid{0000-0001-6280-3851},
H.~Afsharnia$^{9}$,
C.~Agapopoulou$^{13}$\lhcborcid{0000-0002-2368-0147},
C.A.~Aidala$^{78}$\lhcborcid{0000-0001-9540-4988}
% ...

{\footnotesize \it

% ...
$^{9}$Universit{\'e} Clermont Auvergne, CNRS/IN2P3, LPC, Clermont-Ferrand, France\\
% ...
$^{13}$LPNHE, Sorbonne Universit{\'e}, Paris Diderot Sorbonne Paris Cit{\'e}, CNRS/IN2P3, Paris, France\\
% ...
$^{32}$Nikhef National Institute for Subatomic Physics, Amsterdam, Netherlands\\
% ...
$^{41}$Instituto Galego de F{\'\i}sica de Altas Enerx{\'\i}as (IGFAE), Universidade de Santiago de Compostela, Santiago de Compostela, Spain\\
% ...
$^{45}$Physik-Institut, Universit{\"a}t Z{\"u}rich, Z{\"u}rich, Switzerland\\
% ...
$^{49}$H.H. Wills Physics Laboratory, University of Bristol, Bristol, United Kingdom\\
% ...
$^{51}$Department of Physics, University of Warwick, Coventry, United Kingdom\\
% ...
$^{54}$School of Physics and Astronomy, University of Glasgow, Glasgow, United Kingdom\\
$^{55}$Oliver Lodge Laboratory, University of Liverpool, Liverpool, United Kingdom\\
% ...
$^{63}$Syracuse University, Syracuse, NY, United States\\
% ...
$^{77}$Department of Physics and Astronomy, Uppsala University, Uppsala, Sweden, associated to $^{54}$\\
$^{78}$University of Michigan, Ann Arbor, MI, United States, associated to $^{63}$\\
\bigskip
% ...
}
\end{flushleft}

\end{document}